\documentclass[prb,twocolumn,aps,showpacs]{revtex4}
\usepackage{graphicx}%
\usepackage{dcolumn}
\usepackage{amsmath}
\usepackage{amssymb}
\usepackage{epsfig}

\begin{document}

\title{Bethe-Salpeter approach to the collective-mode spectrum of a superfluid
Fermi gas in a moving optical lattice}
\author{Zlatko Koinov,$^1$ Petar Koynov,$^2$}
\affiliation{$^1$Department of Physics and Astronomy, University of
Texas at San Antonio, San Antonio, TX 78249,
USA\\
$^2$Austin Community College, Austin, TX.}
\email{Zlatko.Koinov@utsa.edu} \pacs{03.75.Ss,67.85.-d, 73.20.Mf,
71.10.Pm}
\begin{abstract}
We have derived the Bethe-Salpeter (BS) equations for the
collective-mode spectrum of a  mixture of fermion atoms of two
hyperfine states loaded into a moving optical lattice. The
collective excitation spectrum exhibits rotonlike minimum and the
Landau instability takes place when the energy of the rotonlike
minimum hits zero. The BS approach is compared with the other
existing methods for calculating the collective-mode spectrum. In
particular, it is shown that the spectrum obtained by the BS
equations in an excellent agreement with corresponding spectrum
obtained by the perturbation theory, while the Green's function
formalism provides slightly different results for the collective
excitations.
\end{abstract}
\maketitle

\section{Introduction}

Ultracold atoms loaded in optical lattices are ideal systems to
simulate and study many different problems of correlated quantum
systems because near the Feshbach resonance the atom-atom
interaction can be manipulated in a controllable way by changing the
scattering length $a_s$ from the Bardeen-Cooper-Schrieffer  (BCS)
side (negative values) to the BEC side (positive values) reaching
very large values close to resonance. On Bose-Einstein condensation
(BEC) side of the resonance the spin-up and spin-down atoms can form
diatomic molecules, and these bosonic molecules can undergo a BEC at
low enough temperature. \cite{BEC1,BEC2,BEC3,BEC4} We focus our
attention on the BCS side where BCS superfluidity is expected
analogous to superconductivity.

 In what follows we examine the spectrum of the single-particle
excitations and the spectrum of the collective excitations of a
balanced  mixture of atomic Fermi gas of two hyperfine states with
contact interaction loaded into an optical lattice. The two
hyperfine states are described by pseudospins
$\sigma=\uparrow,\downarrow$. There are $M$ atoms distributed along
$N$ sites, and the corresponding filling factors $f=M/N$ are assumed
to be smaller than unity. If the lattice potential is sufficiently
deep such that the tight-binding approximation is valid, the system
is well described by the single-band attractive Hubbard model:
\begin{equation}H=-J\sum_{<i,j>,\sigma}\psi^\dag_{i,\sigma}\psi_{j,\sigma}
-\mu\sum_{i,\sigma}\widehat{n}_{i,\sigma}+U\sum_i
\widehat{n}_{i,\uparrow} \widehat{n}_{i,\downarrow}.
\label{Hubb1}\end{equation} Here, the Fermi operator
$\psi^\dag_{i,\sigma}$ ($\psi_{i,\sigma}$) creates (destroys) a
fermion on the lattice site $i$ with pseudospin $\sigma$ and
$\widehat{n}_{i,\sigma}=\psi^\dag_{i,\sigma}\psi_{i,\sigma}$ is the
density operator on site $i$ with a position vector $\textbf{r}_i$.
$\mu$ is the corresponding chemical potential, and the symbol
$\sum_{<ij>}$ means sum over nearest-neighbor sites. $J$ is the
tunneling strength of the atoms between nearest-neighbor sites, and
$U$ is the on-site interaction. On the BCS side the interaction
parameter $U$ is negative (the atomic interaction is attractive). We
assume $\hbar=1$, Boltzmann constant $k_B=1$, $J=1$ and the lattice
constant $a=1$.

In the case when the periodic array of microtraps is generated by
counter propagating laser beams with different frequencies the
optical lattice potential is moving with a velocity $-\textbf{v}$
(in the laboratory frame) which magnitude is proportional to the
relative frequency detuning of the two laser beams. In a frame fixed
with respect to the lattice potential the fermion atoms flows with a
constant quasimomentum $\textbf{p}=m\textbf{v}$,  where $m$ is the
mass of the loaded atoms.  For balance Fermi gases the order
parameter field
$\Phi_j(u)=-|U|<\psi_{j,\downarrow}(u)\psi_{j,\uparrow}(u)>$ (or
$\Phi^*_j(u)=-|U|<\psi^\dag_{j,\uparrow}(u)\psi^\dag_{j,\downarrow}(u)>$)
in the mean-field approximation varies as
$\Phi_j\propto\Delta_0\exp\left[2\imath\textbf{p.}\textbf{r}_j\right]$.
Here, the symbol $< >$ means ensemble average, and $\Delta$ is a
real quantity which depends on the lattice velocity.\cite{Gam, Yosh}
In a moving lattice the formation of a BCS superfluidity is
possible, but due to the presence of quasimomentum $\textbf{p}$ the
superflow can break down. The stability of balance superfluid Fermi
gases loaded into a moving optical lattice has been recently studied
using the second-order time-dependent perturbation theory
\cite{Gam}, and the Green's function formalism. \cite{Yosh}  The
superfluid state could be destabilized at a critical flow momentum
via two different mechanisms: depairing (pair-breaking) at
$\textbf{p}_{pb}$ and Landau instabilities at $\textbf{p}_{cr}$. The
depairing takes place when the single fermionic excitations are
broken, while the Landau instability is related to the rotonlike
structure of the spectrum of the collective excitations. The
superfluid state becomes unstable when the energy of the rotonlike
minimum reaches zero at a given quasimomentum.  The numerical
solution of the number, gap and the collective-mode equations shows
that at a zero temperature the Landau instability appears before the
depearing mechanism.\cite{Yosh}

Generally speaking, there exist three methods that can be used to
calculate the spectrum of the collective excitations of Hamiltonian
(\ref{Hubb1}) in a stationary (or moving) optical lattice in
generalized random phase approximation (GRPA). The first approach
uses the Green's function
method,\cite{CGexc,CCexc,CG1,ZKexc,Com,ZGK,ZK1}  the second one is
based on the Anderson-Rickayzen equations,\cite{PA,R,BR} while the
third employs the perturbation technique.\cite{Gam}

Decades ago, the Green's function approach has been used to obtain
the collective excitations in the exciton problem \cite{CGexc,CCexc}
and in the s-wave layered superconductors. \cite{CG1} In both cases
we have deal with a system of interacting electrons, where two
different interactions exist: the Coulomb interaction
$v(\textbf{r})=e^2/|\textbf{r}|$ and a short-range attractive
interaction $g(\textbf{r})$ between electrons. In the GRPA the
collective-mode spectrum can be  obtained by the poles of: (i) the
vertex $\Gamma$  from the corresponding BS equation for $\Gamma$,
\cite{CCexc,Com} (ii) the two-particle Green's function $K$ by
solving the BS equation for the BS amplitudes, \cite{ZKexc,ZGK,ZK1}
and (iii) the density and spin response functions.\cite{CGexc,CG1}
It is worth mentioning that in the exciton problem the BS vertex
equation \cite{CCexc} and the BS equation
$K^{-1}=K^{(0)-1}-I_d-I_{exc}$ (here $K^{(0)}$ is the free
two-particle propagator, and $I_d$ takes into account the direct
interaction between the electrons, while $I_{exc}$ describes their
exchange interaction\cite{ZKexc}), both provide the same
collective-mode spectrum, while in the case of superconductivity,
the vertex equation obtained  in Ref. [\onlinecite{Com}]  in for the
collective-mode dispersion (see Eq. (14) in Ref. [\onlinecite{Com}])
is incorrect because the exchange interaction has been neglected
(the incorrect equation follows from our secular determinant
(\ref{BSZ}) assuming a stationary lattice and neglecting all
elements with $\widetilde{\gamma}$ and $m$ indices).

C\^{o}t\'{e} and Griffin \cite{CGexc,CG1} have obtained the
collective excitation spectrum from the poles of the density and
spin response functions using the Baym and Kadanoff formalism.
\cite{BK} They have ignored the long-range Coulomb interaction
keeping in the direct interaction only the ladder diagrams involving
the short-range interaction. The exchange interaction involves
bubble diagrams with respect to both the unscreened Coulomb
interaction $v(\textbf{r})$ and the short-range attractive
interaction $g(\textbf{r})$. Thus, Eq. (2.33) in Ref.
[\onlinecite{CG1}] corresponds to the BS equation
$\widetilde{K}^{-1}=K^{(0)-1}-I_d$, and Eq. (2.32) in Ref.
 [\onlinecite{CG1}] corresponds to the BS equation
$K=\widetilde{K}+\widetilde{K}I_{exc}K$. The C\^{o}t\'{e} and
Griffin Green's function formalism was employed in Ref.
[\onlinecite{Yosh}] to study the collective-mode spectrum in moving
optical lattices by obtaining the poles of the density response
function. In what follows we will see that the BS approach provides
collective-mode spectrum which lies slightly below the one obtained
 in Ref. [\onlinecite{Yosh}] if the both methods use the same gap and chemical potential.
  The most important advantage of the BS
formalism over the Green's function approach is that within the BS
approach we can obtain the poles of density and spin response
functions in a uniform manner, i.e. one secular determinant provides
not only the poles of the density response function, but the poles
of the spin response function as well.

The second method \cite{BR} that can be used to obtain the
collective excitation spectrum of Hamiltonian (\ref{Hubb1}) starts
from the Anderson-Rickayzen equations, which in the GRPA were
reduced to a set of three coupled equations and the collective-mode
spectrum is obtained by solving the secular equation $Det|A|=0$,
where
\begin{equation}
A=\left(
\begin{array}{ccc}
|U|^{-1}+I_{l,l}&J_{\gamma,l}
&I_{l,m}\\
I_{\gamma,l}&|U|^{-1}+I_{\gamma,\gamma}
&J_{\gamma,m}\\
I_{l,m}&J_{\gamma,m} &|U|^{-1}+ I_{m,m}
\end{array}%
\right).\label{Z1}
\end{equation} In a stationary lattice, the quantities $I$ and $J$ are
defined in Sec. II by Eqs. (\ref{JSC}) when $\textbf{p}=0$. As can
be seen, the BS secular determinant provides an equation for the
collective-mode dispersion that depends on all four coherence
factors and the associated  determinant is $4\times 4$, while the
determinant (\ref{Z1}) does not depend on the fourth coherence
factor $\widetilde{\gamma}$, so that the corresponding determinant
is only $3\times 3$.

Recently, Ganesh et al.\cite{Gam} have used a perturbation approach
 which in the case of a moving optical lattice
 provides the following $3\times 3$ secular determinant (see Eqs.
(B8) in Ref. [\onlinecite{Gam}], where there is a negative sign in
front of the sum in the definition of
$\chi_0^{2,3}$):\begin{widetext}
\begin{equation} D=\left(
\begin{array}{ccc}
|U|^{-1}+I^\textbf{p}_{m,m}&(J^\textbf{p}_{\gamma,m}-I^\textbf{p}_{lm})/2
&-(J^\textbf{p}_{\gamma,m}+I^\textbf{p}_{l,m})/2\\
I^\textbf{p}_{l,m}-J^\textbf{p}_{\gamma,
m}&|U|^{-1}+(I^\textbf{q}_{l,l}+
I^\textbf{p}_{\gamma,\gamma}-2J^\textbf{p}_{\gamma,
l})/2&(I^\textbf{p}_{\gamma,\gamma}
-I^\textbf{p}_{l,l})/2\\
-I^\textbf{p}_{l,m}-J^\textbf{p}_{\gamma,m}&(I^\textbf{p}_{\gamma,\gamma}
-I^\textbf{p}_{l,l})/2&|U|^{-1}+
(I^\textbf{p}_{l,l}+I^\textbf{p}_{\gamma,\gamma}+2J^\textbf{p}_{\gamma,
l})/2
\end{array}%
\right). \label{Gam}\end{equation}\end{widetext}  Since
$Det|D|=det|A^{\textbf{p}}|$, where
\begin{equation} A^{\textbf{p}}=\left|
\begin{array}{cccc}
|U|^{-1}+I^\textbf{p}_{\gamma,\gamma}&J^\textbf{p}_{\gamma,l}
&J^\textbf{p}_{\gamma,m}\\
J^\textbf{p}_{\gamma,l}&|U|^{-1}+I^\textbf{p}_{l,l}&I^\textbf{p}_{l,m}
\\
J^\textbf{p}_{\gamma,m}
&I^\textbf{p}_{l,m}&|U|^{-1}+I^\textbf{p}_{m,m}\end{array}\right|,
\label{Josh1}\end{equation} one can say that the perturbation method
is a generalization of the Belkhir and Randeria method to the case
of moving optical lattice.

This paper is organized as follows. In Sec. II, we derive the BS
equations for the dispersion of the collective excitations of
imbalanced  mixture of fermionic atoms loaded in a moving optical
lattice. In Sec. III, we compare our approach with the other
existing methods, and we discuss the stability of balance gases
loaded in a moving lattice. It turns out that the BS approach and
the second-order time-dependent perturbation theory\cite{Gam}
provide essentially the same spectrum of the collective excitations,
while the spectrum obtained by the Green's function
method\cite{Yosh} differs from the corresponding dispersion obtained
by the BS approach. But, the Green's function method and the BS
approach, both predict that as the quasimomentum $\textbf{p}$
increases, the rotonlike minimum reaches zero before the
pair-breaking mechanism takes place.
\section{Bethe-Salpeter equation for the spectrum of the collective modes}
A basic approximation in our approach is that the single-particle
excitations in the system will be calculated in the mean-field
approximation, i.e. our single-particle Green's function in the
momentum and frequency space is the following $2\times 2$ matrix
$\widehat{G}=\left(
\begin{array}{cc}
G^{\uparrow,\uparrow}&G^{\uparrow,\downarrow}\\
G^{\downarrow,\uparrow} &G^{\downarrow,\downarrow}
\end{array}%
\right)$. The the corresponding matrix elements are defined as
follows:
$$
G^{ij}_{\textbf{p}}(\textbf{k},\imath\omega_m)=
\frac{A^{ij}_\textbf{p}(\textbf{k})}
{\imath\omega_m-E_{+}(\textbf{k};\textbf{p})}+
\frac{B^{ij}_\textbf{p}(\textbf{k})}
{\imath\omega_m-E_{-}(\textbf{k};\textbf{p})},$$
where\begin{equation}\begin{split}&
\widehat{A}_\textbf{p}(\textbf{k})=\left(
\begin{array}{cc}
u^2_{\textbf{p}}(\textbf{k})&u_{\textbf{p}}(\textbf{k})v_{\textbf{p}}(\textbf{k})\\
u_{\textbf{p}}(\textbf{k})v_{\textbf{p}}(\textbf{k})
&v^2_{\textbf{p}}(\textbf{k})
\end{array}%
\right) \nonumber\\& \widehat{B}_\textbf{p}(\textbf{k})=\left(
\begin{array}{cc}
v^2_{\textbf{p}}(\textbf{k})&-u_{\textbf{p}}(\textbf{k})v_{\textbf{p}}(\textbf{k})\\
-u_{\textbf{p}}(\textbf{k})v_{\textbf{p}}(\textbf{k})
&u^2_{\textbf{p}}(\textbf{k})
\end{array}%
\right). \nonumber\end{split}\end{equation} The symbol $\omega_m$
denotes $\omega_m= (2\pi/\beta)(m +1/2) ; \beta=T^{-1}$, $T$ is the
temperature,  and $m=0, \pm 1, \pm 2,... $. Here,
$u_{\textbf{p}}(\textbf{k})=\sqrt{\frac{1}{2}\left[1+
\chi(\textbf{k};\textbf{p})/E_{\textbf{p}}(\textbf{k})\right]}$ and
$v_{\textbf{p}}(\textbf{k})=\sqrt{\frac{1}{2}\left[1-
\chi(\textbf{k};\textbf{p})/E_{\textbf{p}}(\textbf{k})\right]}$ are
the coherent factors, and we have introduced the following notations
$\chi(\textbf{k};\textbf{p})=
\frac{1}{2}\left[\xi(\textbf{p}+\textbf{k})+
\xi(\textbf{k}-\textbf{p})\right]$, $\eta(\textbf{k};\textbf{p})=
\frac{1}{2}\left[\xi(\textbf{p}+\textbf{k})-
\xi(\textbf{k}-\textbf{p})\right]$, $E_{\pm}(\textbf{k};\textbf{p})=
\eta(\textbf{k};\textbf{p})\pm E_{\textbf{p}}(\textbf{k})$, and
$E_{\textbf{p}}(\textbf{k})=
\sqrt{\chi^2(\textbf{k};\textbf{p})+\Delta^2}$.  The tight-binding
form of the electron energy is $\xi(\textbf{k})=2J\left(1-\sum_\nu
\cos k_\nu \right)-\mu$. As can be seen, the single-particle
excitations in the mean-field approximation are coherent
combinations of electronlike $E_{+}$ and holelike $E_{-}$ fermionic
excitations.

 The thermodynamic potential
in a mean-field approximation can be evaluated as a summation of
quasiparticles with energy $E_\pm(\textbf{k})$:
\begin{equation}\begin{split}&
\Omega=
\frac{\Delta^2}{|U|}+\frac{1}{N}\sum_\textbf{k}\{\chi(\textbf{k};\textbf{p})
-E_{\textbf{p}}(\textbf{k})\\&-\frac{1}{\beta}\ln\left(\left[1+\exp\left(-\beta
E_{+}(\textbf{k};\textbf{p})\right)\right]\left[1+\exp\left(\beta
E_{-}(\textbf{k};\textbf{p})\right)\right]\right)\}.
 \nonumber\end{split}\end{equation}  In the case of balance
population  of fermionic atoms the gap and particle number equations
can be obtained by  minimizing the free energy $F(\Delta,f)=\Omega+
\mu f$ with respect to $\mu$ and $\Delta$:
\begin{equation}\begin{split}&
f=\frac{2}{N}\sum_\textbf{k} \left[u_{\textbf{p}}^2(\textbf{k})
f(E_{+}(\textbf{k};\textbf{p})+
v_{\textbf{p}}^2(\textbf{k})f(E_{-}(\textbf{k};\textbf{p})\right],
\\& 1=\frac{|U|}{N}\sum_\textbf{k}
\frac{f(E_{-}(\textbf{k};\textbf{p}))-
f(E_{+}(\textbf{k};\textbf{p}))}{2E_{\textbf{p}}(\textbf{k})},
\label{MO}\end{split}\end{equation} where
$f(x)=\left[\exp\left(\beta x\right)+1\right]^{-1}$ is the Fermi
distribution function.
\begin{figure}
    \includegraphics[scale=0.99]{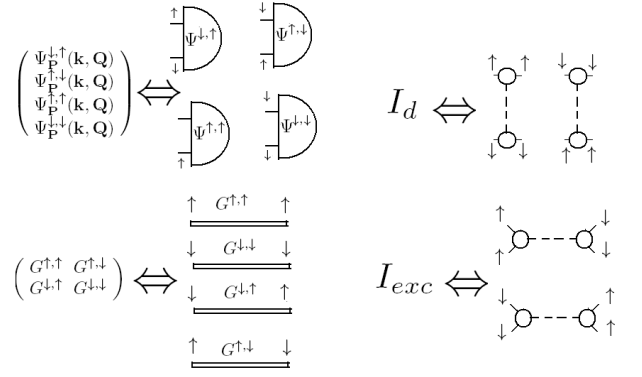}
    \caption{Diagrammatic representation of the Bethe-Salpeter amplitude, the single-particle
     Green's function, the direct and exchange interactions.}
\end{figure}

\begin{figure}
    \includegraphics[scale=0.99]{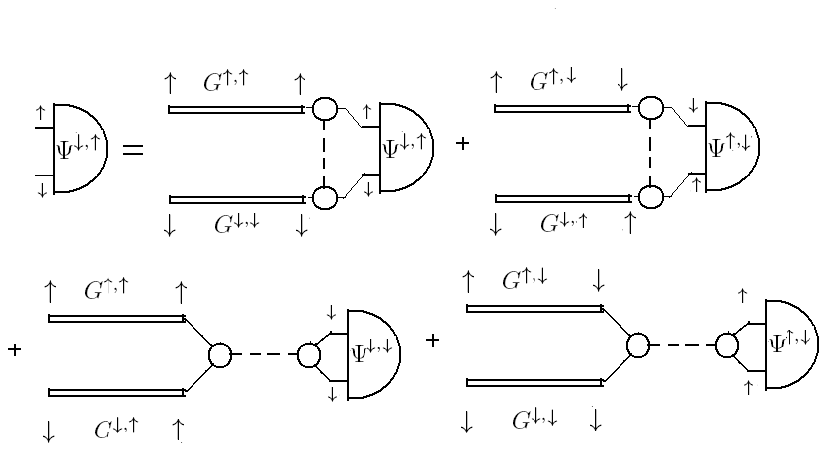}
    \caption{ Diagrammatic representation of the Bethe-Salpeter equation for the amplitude
    $\Psi_\textbf{p}^{\downarrow,\uparrow}(\textbf{k},\textbf{Q})$. }
\end{figure}
Next, we derive the BS equations for the spectrum of the collective
modes in a moving lattice. The collective mode energies
$\omega_\textbf{p}(\textbf{Q})$ and the corresponding BS amplitude
$\widehat{\Psi}_\textbf{p}(\textbf{k},\textbf{Q})=\left(
\begin{array}{c}
\Psi_\textbf{p}^{\downarrow,\uparrow}(\textbf{k},\textbf{Q})\\
\Psi_\textbf{p}^{\uparrow,\downarrow}(\textbf{k},\textbf{Q}) \\
\Psi_\textbf{p}^{\uparrow,\uparrow}(\textbf{k},\textbf{Q})\\
\Psi_\textbf{p}^{\downarrow,\downarrow}(\textbf{k},\textbf{Q})\\
\end{array}%
\right)$ in a moving optical lattice and spin-polarized systems can
be derived in a similar manner as in a stationary lattice. Our
approach is based on the fact that the Hubbard model with on-site
attractive interaction can be transformed to a model in which the
narrow-band free electrons are coupled to a boson field due to some
spin-dependent mechanism.\cite{ZGK,ZK1} We first use the
Hubbard-Stratonovich transformation to transform the quartic Hubbard
term in (\ref{Hubb1}) to a quadratic form. Thus,  the quartic
problem of interacting electrons is transformed to a quadratic
problem of non-interacting Nambu fermion fields coupled to a Bose
field and therefore, we arrive at the problem of the linear response
of many-fermion systems under a weak bosonic field when a transition
to a quantum condensed phase is possible. We have similar problem
 in exciton physics when the BEC of excitons can be described
by applying the powerful arsenal of quantum field theory to obtain
exact equations for the single-particle Green's function, mass
operator  and the BS equation for the two-particle Green's
function.\cite{CGexc,CCexc,ZKexc} As in the exciton problem, the
mass operator is a sum of two terms: the Fock $\Sigma_F$ and Hartree
$\Sigma_H$ terms. The Hartree term is diagonal with respect to the
spin indices and it generates the exchange interaction
$I_{exc}=\delta\Sigma_H/\delta G$ in the BS equation for the
spectrum of the collective excitations, while the direct interaction
$I_{d}=\delta\Sigma_F/\delta G$ in the kernel of the BS equation
originates from the Fock term. In Fig. 1 we have shown the
diagrammatic representations of the leading terms of the direct and
exchange interactions. In this approximation the BS equation for the
BS amplitude is $\Psi=K^{(0)}\left[I_{d}+I_{exc}\right]\Psi$, where
$K^{(0)}$ is a product of two single-particle Green's functions in
the mean-field approximation. In Fig. 2 we have shown the BS
equation for
$\Psi_\textbf{p}^{\downarrow,\uparrow}(\textbf{k},\textbf{Q})$. The
other three equations for
$\Psi_\textbf{p}^{\uparrow,\downarrow}(\textbf{k},\textbf{Q})$, $
\Psi_\textbf{p}^{\uparrow,\uparrow}(\textbf{k},\textbf{Q})$ and $
\Psi_\textbf{p}^{\downarrow,\downarrow}(\textbf{k},\textbf{Q})$ are
similar to that one. We can write all four BS equations as a single
matrix equation: \begin{widetext}
\begin{equation}
\widehat{\Psi}_\textbf{p}(\textbf{k},\textbf{Q})=
-\frac{|U|}{2N}\widehat{D}_\textbf{p}(\textbf{k},\textbf{Q},\imath\omega_p)\sum_\textbf{q}
\widehat{\Psi}_\textbf{p}(\textbf{q},\textbf{Q})+\frac{|U|}{2N}
\widehat{M}_\textbf{p}(\textbf{k},\textbf{Q},\imath\omega_p)\sum_\textbf{q}
\widehat{\Psi}_\textbf{p}(\textbf{q},\textbf{Q}).
\label{BS}\end{equation}

Here,
\begin{equation}
\widehat{D}=\left(
\begin{array}{cccc}
K_\textbf{p}^{\left(\downarrow,\downarrow,\uparrow,\uparrow\right)}(\textbf{k},\textbf{Q},\imath\omega_p),&
K_\textbf{p}^{\left(\downarrow,\uparrow,\downarrow,\uparrow\right)}(\textbf{k},\textbf{Q},\imath\omega_p)&0&0\\
K_\textbf{p}^{\left(\uparrow,\downarrow,\uparrow,\downarrow\right)}(\textbf{k},\textbf{Q},\imath\omega_p),&
K_\textbf{p}^{\left(\uparrow,\uparrow,\downarrow,\downarrow\right)}(\textbf{k},\textbf{Q},\imath\omega_p)&0&0\\
K_\textbf{p}^{\left(\uparrow,\downarrow,\uparrow,\uparrow\right)}(\textbf{k},\textbf{Q},\imath\omega_p),&
K_\textbf{p}^{\left(\uparrow,\uparrow,\downarrow,\uparrow\right)}(\textbf{k},\textbf{Q},\imath\omega_p)&0&0\\
K_\textbf{p}^{\left(\downarrow,\downarrow,\uparrow,\downarrow\right)}(\textbf{k},\textbf{Q},\imath\omega_p),&
K_\textbf{p}^{\left(\downarrow,\uparrow,\downarrow,\downarrow\right)}(\textbf{k},\textbf{Q},\imath\omega_p)&0&0
\end{array}%
\right),\widehat{M}=\left(
\begin{array}{cccc}
0&0&K_\textbf{p}^{\left(\downarrow,\downarrow,\downarrow,\uparrow\right)}(\textbf{k},\textbf{Q},\imath\omega_p),&
K_\textbf{p}^{\left(\downarrow,\uparrow,\uparrow,\uparrow\right)}(\textbf{k},\textbf{Q},\imath\omega_p)\\
0&0&K_\textbf{p}^{\left(\uparrow,\downarrow,\downarrow,\downarrow\right)}(\textbf{k},\textbf{Q},\imath\omega_p),&
K_\textbf{p}^{\left(\uparrow,\uparrow,\uparrow,\downarrow\right)}(\textbf{k},\textbf{Q},\imath\omega_p)\\
0&0&K_\textbf{p}^{\left(\uparrow,\downarrow,\downarrow,\uparrow\right)}(\textbf{k},\textbf{Q},\imath\omega_p),&
K_\textbf{p}^{\left(\uparrow,\uparrow,\uparrow,\uparrow\right)}(\textbf{k},\textbf{Q},\imath\omega_p)\\
0&0&K_\textbf{p}^{\left(\downarrow,\downarrow,\downarrow,\downarrow\right)}(\textbf{k},\textbf{Q},\imath\omega_p),&
K_\textbf{p}^{\left(\downarrow,\uparrow,\uparrow,\downarrow\right)}(\textbf{k},\textbf{Q},\imath\omega_p)
\end{array}%
\right),
 \nonumber\end{equation}
 and the terms  $|U|\widehat{D}=K^{(0)}I_d$ and
$|U|\widehat{M}=K^{(0)}I_{exc}$ represent the contributions due to
the direct and exchange interactions, respectively. The two-particle
propagator
$K_\textbf{p}^{\left(i,j,k,l\right)}(\textbf{k},\textbf{Q},\imath\omega_p)$
is:
\begin{equation}\begin{split}
&K_\textbf{p}^{\left(i,j,k,l\right)}(\textbf{k},\textbf{Q},\imath\omega_p\rightarrow
\omega+\imath0^+)=\sum_{\omega_m}G_\textbf{p}^{i,j}(\textbf{k}+\textbf{Q};\imath\omega_p+\imath\omega_m)
G_\textbf{p}^{k,l}(\textbf{k};\imath\omega_m)|_{\imath\omega_p\rightarrow
\omega+\imath0^+}=\\& \widetilde{A}^{ij}_\textbf{p}A^{kl}_\textbf{p}
\frac{f(E_+(\textbf{k};\textbf{p}))-f(E_+(\textbf{k}+\textbf{Q};\textbf{p}))}{\omega-\left[
E_+(\textbf{k}+\textbf{Q};\textbf{p})-E_+(\textbf{k};\textbf{p})\right]+\imath0^+}
+\widetilde{B}^{ij}_\textbf{p}B^{kl}_\textbf{p}
\frac{f(E_-(\textbf{k};\textbf{p}))-f(E_-(\textbf{k}+\textbf{Q};\textbf{p}))}{\omega-\left[
E_-(\textbf{k}+\textbf{Q};\textbf{p})-E_-(\textbf{k};\textbf{p})\right]+\imath0^+}\\&
+\widetilde{A}^{ij}_\textbf{p}B^{kl}_\textbf{p}
\frac{f(E_-(\textbf{k};\textbf{p}))-f(E_+(\textbf{k}+\textbf{Q};\textbf{p}))}{\omega-\left[
E_+(\textbf{k}+\textbf{Q};\textbf{p})-E_-(\textbf{k};\textbf{p})\right]+\imath0^+}
+A^{ij}_\textbf{p} \widetilde{B}^{kl}_\textbf{p}
\frac{f(E_+(\textbf{k};\textbf{p}))-f(E_-(\textbf{k}+\textbf{Q};\textbf{p}))}{\omega-\left[
E_-(\textbf{k}+\textbf{Q};\textbf{p})-E_+(\textbf{k};\textbf{p})\right]+\imath0^+}
\end{split}\end{equation}
Here $ i,j,k,l=\{\uparrow,\downarrow\}$ and $\omega_{p}=(2\pi/
\beta)p ; p=0, \pm 1, \pm 2,...$ is a Bose frequency,
$A^{ij}_\textbf{p}=A^{ij}_\textbf{p}(\textbf{k})$,
$B^{ij}_\textbf{p}=B^{ij}_\textbf{p}(\textbf{k})$ ,
$\widetilde{A}^{ij}_\textbf{p}=A^{ij}_\textbf{p}(\textbf{k}+\textbf{Q})$
and
$\widetilde{B}^{ij}_\textbf{p}=B^{ij}_\textbf{p}(\textbf{k}+\textbf{Q})$.
To solve the BS equations we introduce a new matrix
$\widehat{\Phi}_\textbf{p}(\textbf{k},\textbf{Q})=
\widehat{T}\widehat{\Psi}_\textbf{p}(\textbf{k},\textbf{Q})$, where
$ \widehat{T}=\left(\begin{array}{cc}
\sigma_x+\sigma_z&0\\
0&\sigma_x+\sigma_z\\
\end{array}%
\right)$ and $\sigma_x$ and $\sigma_z$ are the Pauli matrices. By
means of $\widehat{\Phi}_\textbf{p}(\textbf{k},\textbf{Q})$ we
obtain the following equations for collective modes:
\begin{equation}
\frac{1}{N}\sum_\textbf{k}\widehat{\Phi}_\textbf{p}(\textbf{k},\textbf{Q})
=-\frac{|U|}{2N}\sum_\textbf{k}\widehat{T}\left[
\widehat{D}_\textbf{p}(\textbf{k},\textbf{Q},\imath\omega_p
)-\widehat{M}_\textbf{p}(\textbf{k},\textbf{Q},\imath\omega_p)\right]\widehat{T}^{-1}
\frac{1}{N}\sum_\textbf{q}\widehat{\Phi}_\textbf{p}(\textbf{q},\textbf{Q}).
\nonumber\end{equation} The condition for existing a non-trivial
solution  leads to the vanishing of the following secular
determinant:
\begin{equation}
Z= \left|
\begin{array}{cccc}
|U|^{-1}+\left(I_{\gamma,\gamma}-L_{\widetilde{\gamma},\widetilde{\gamma}}\right)&\left(J_{\gamma,l}-K_{m,\widetilde{\gamma}}\right)&
\left(I_{\gamma,\widetilde{\gamma}}+L_{\gamma,\widetilde{\gamma}}\right)&\left(J_{\gamma,m}+K_{l,\widetilde{\gamma}}\right)\\
\left(J_{\gamma,l}-K_{m,\widetilde{\gamma}}\right)&|U|^{-1}+\left(I_{l,l}-L_{m,m}\right)&
\left(J_{l,\widetilde{\gamma}}+K_{m,\gamma}\right)&\left(I_{l,m}+L_{l,m}\right)\\
\left(I_{\gamma,\widetilde{\gamma}}+L_{\gamma,\widetilde{\gamma}}\right)&\left(J_{l,\widetilde{\gamma}}+K_{m,\gamma}\right)&
-|U|^{-1}+\left(I_{\widetilde{\gamma},\widetilde{\gamma}}-L_{\gamma,\gamma}\right)&\left(J_{\widetilde{\gamma},m}-K_{\gamma,l}\right)\\
\left(J_{\gamma,m}+K_{l,\widetilde{\gamma}}\right)&\left(I_{l,m}+L_{l,m}\right)&
\left(J_{\widetilde{\gamma},m}-K_{\gamma,l}\right)&|U|^{-1}+\left(I_{m,m}-L_{l,l}\right)
\end{array}%
\right|,\label{SecDet}\end{equation} where the following symbols
have been introduced:
\begin{equation}\begin{split}&
I_{a,b}=\frac{1}{2N}\sum_\textbf{k}a^{\textbf{p}}_{\textbf{k},\textbf{Q}}
b^{\textbf{p}}_{\textbf{k},\textbf{Q}}
\left[\frac{f\left(E_{-}(\textbf{k};\textbf{p})
\right)-f\left(E_{+}(\textbf{k}+\textbf{Q};\textbf{p})
\right)}{\omega+\Omega_{\textbf{p}}(\textbf{k},\textbf{Q})-
\varepsilon_{\textbf{p}}(\textbf{k},\textbf{Q})}
-\frac{f\left(E_{-}(\textbf{k}+\textbf{Q};\textbf{p})
\right)-f\left(E_{+}(\textbf{k};\textbf{p})
\right)}{\omega+\Omega_{\textbf{p}}(\textbf{k},
\textbf{Q})+\varepsilon_{\textbf{p}}(\textbf{k},\textbf{Q})}\right]
,\nonumber\\& J_{a,b}=\frac{1}{2N}\sum_\textbf{k}
a^{\textbf{p}}_{\textbf{k},\textbf{Q}}
b^{\textbf{p}}_{\textbf{k},\textbf{Q}}
\left[\frac{f\left(E_{-}(\textbf{k};\textbf{p})
\right)-f\left(E_{+}(\textbf{k}+\textbf{Q};\textbf{p}) \right)}
{\omega+\Omega_{\textbf{p}}(\textbf{k},\textbf{Q})-
\varepsilon_{\textbf{p}}(\textbf{k},\textbf{Q})}
+\frac{f\left(E_{-}(\textbf{k}+\textbf{Q};\textbf{p})
\right)f\left(E_{+}(\textbf{k};\textbf{p}) \right)}
{\omega+\Omega_{\textbf{p}}(\textbf{k},
\textbf{Q})+\varepsilon_{\textbf{p}}(\textbf{k},\textbf{Q})}\right]
,\nonumber\\& K_{a,b}=\frac{1}{2N}\sum_\textbf{k}
a^{\textbf{p}}_{\textbf{k},\textbf{Q}}
b^{\textbf{p}}_{\textbf{k},\textbf{Q}}
\left[\frac{f\left(E_{-}(\textbf{k}+\textbf{Q};\textbf{p})
\right)-f\left(E_{-}(\textbf{k};\textbf{p}) \right)}
{\omega+\Omega_{\textbf{p}}(\textbf{k},\textbf{Q})+
\epsilon_{\textbf{p}}(\textbf{k},\textbf{Q})}
-\frac{f\left(E_{+}(\textbf{k}+\textbf{Q};\textbf{p})
\right)-f\left(E_{+}(\textbf{k};\textbf{p}) \right)}
{\omega+\Omega_{\textbf{p}}(\textbf{k},\textbf{Q})-
\epsilon_{\textbf{p}}(\textbf{k},\textbf{Q})}\right] ,\nonumber\\&
L_{a,b}=\frac{1}{2N}\sum_\textbf{k}
a^{\textbf{p}}_{\textbf{k},\textbf{Q}}
b^{\textbf{p}}_{\textbf{k},\textbf{Q}}
\left[\frac{f\left(E_{-}(\textbf{k}+\textbf{Q};\textbf{p})
\right)-f\left(E_{-}(\textbf{k};\textbf{p}) \right)}
{\omega+\Omega_{\textbf{p}}(\textbf{k},\textbf{Q})+
\epsilon_{\textbf{p}}(\textbf{k},\textbf{Q})}
+\frac{f\left(E_{+}(\textbf{k}+\textbf{Q};\textbf{p})
\right)-f\left(E_{+}(\textbf{k};\textbf{p}) \right)}
{\omega+\Omega_{\textbf{p}}(\textbf{k},\textbf{Q})-
\epsilon_{\textbf{p}}(\textbf{k},\textbf{Q})}\right]
.\label{ME}\end{split}\end{equation}\end{widetext}Here,
$\varepsilon_{\textbf{p}}(\textbf{k},\textbf{Q})=
 E_{\textbf{p}}(\textbf{k}+\textbf{Q})+ E_{\textbf{p}}(\textbf{k})$,
  $\epsilon_{\textbf{p}}(\textbf{k},\textbf{Q})=
 E_{\textbf{p}}(\textbf{k}+\textbf{Q})- E_{\textbf{p}}(\textbf{k})$,
  $\Omega_{\textbf{p}}(\textbf{k},\textbf{Q})
  =\eta_{\textbf{p}}(\textbf{k})-\eta_{\textbf{p}}
 (\textbf{k}+\textbf{Q})$,
  and $a$ and $b$ are one of the following form factors:
\begin{equation}\begin{split}&\gamma^{\textbf{p}}_{\textbf{k},\textbf{Q}}=
u_{\textbf{p}}(\textbf{k})u_{\textbf{p}}(\textbf{k}+\textbf{Q})
+v_{\textbf{p}}(\textbf{k})v_{\textbf{p}}(\textbf{k}+\textbf{Q}),
\nonumber\\&
l^{\textbf{p}}_{\textbf{k},\textbf{Q}}=u_{\textbf{p}}(\textbf{k})
u_{\textbf{p}}(\textbf{k}+\textbf{Q})
-v_{\textbf{p}}(\textbf{k})v_{\textbf{p}}(\textbf{k}+\textbf{Q}),\nonumber\\&
\widetilde{\gamma}_{\textbf{p}}{\textbf{k},\textbf{Q}}=
u_{\textbf{p}}(\textbf{k})v_{\textbf{p}}(\textbf{k}+\textbf{Q})
-v_{\textbf{p}}(\textbf{k})u_{\textbf{p}}(\textbf{k}+\textbf{Q}),\nonumber\\&
 m^{\textbf{p}}_{\textbf{k},\textbf{Q}}=
u_{\textbf{p}}(\textbf{k})v_{\textbf{p}}(\textbf{k}+\textbf{Q})
+v_{\textbf{p}}(\textbf{k})u_{\textbf{p}}(\textbf{k}+\textbf{Q}).\nonumber\end{split}\end{equation}
As $\textbf{Q}\rightarrow 0$ in accordance with the well-known
Goldstone theorem, there exists a solution $\omega\rightarrow 0$. In
this case all $J$, $K$ and $L$ vanishes, and the secular equation
reduces to the gap equation written as
$0=1+|U|I_{\gamma=1,\gamma=1}$.

At a zero temperature, before the pair breaking sets in, we have
  $E_{+}(\textbf{k};\textbf{p})>0$,
$E_{-}(\textbf{k}; \textbf{p})=-E_{+}(\textbf{k}; \textbf{p})$, and
the secular
 determinant (\ref{SecDet}) assumes the form

\begin{equation} \left|
\begin{array}{cccc}
|U|^{-1}+I^\textbf{p}_{\gamma,\gamma}&J^\textbf{p}_{\gamma,l}
&I^\textbf{p}_{\gamma,\widetilde{\gamma}}&J^\textbf{q}_{\gamma,m}\\
J^\textbf{p}_{\gamma,l}&|U|^{-1}+I^\textbf{p}_{l,l}
&J^\textbf{p}_{l,\widetilde{\gamma}}&I^\textbf{p}_{l,m}\\
I^\textbf{p}_{\gamma,\widetilde{\gamma}}&J^\textbf{p}_{l,\widetilde{\gamma}}&
-|U|^{-1}+I^\textbf{p}_{\widetilde{\gamma},\widetilde{\gamma}}&
J^\textbf{p}_{\widetilde{\gamma},m}\\
J^\textbf{p}_{\gamma,m}&I^\textbf{p}_{l,m}&
J^\textbf{p}_{\widetilde{\gamma},m}&|U|^{-1}+I^\textbf{p}_{m,m}\end{array}%
\right|. \label{BSZ}\end{equation}  Here we have introduced symbols
$I^\textbf{p}_{a,b}$ and $J^\textbf{p}_{a,b}$:
\begin{equation}\begin{split}&
I^\textbf{p}_{a,b}=\frac{1}{N}\sum_{\textbf{k}}\frac{a^\textbf{p}_{\mathbf{k}%
,\mathbf{Q}}b^\textbf{p}_{\mathbf{k}%
,\mathbf{Q}}\varepsilon_\textbf{p}(\textbf{k},\textbf{Q})}
{\left[\omega+\Omega_\textbf{p}(\textbf{k},\textbf{Q})\right]^2-\varepsilon^2_\textbf{p}(\textbf{k},\textbf{Q})}
,\\&
J^\textbf{p}_{a,b}=\frac{1}{N}\sum_{\textbf{k}}\frac{a^\textbf{p}_{\mathbf{k}%
,\mathbf{Q}}b^\textbf{p}_{\mathbf{k}%
,\mathbf{Q}}\left[\omega+\Omega_\textbf{p}(\textbf{k},\textbf{Q})\right]}
{\left[\omega+\Omega_\textbf{p}(\textbf{k},\textbf{Q})\right]^2
-\varepsilon^2_\textbf{p}(\textbf{k},\textbf{Q})},
\label{JSC}\end{split}\end{equation}
 and the quantities
$a^\textbf{p}_{\mathbf{k}%
,\mathbf{Q}}$ and $b^\textbf{p}_{\mathbf{k}%
,\mathbf{Q}}$ could be one of the
four form factors: $l^\textbf{p}_{\mathbf{k},\mathbf{Q}},m^\textbf{p}_{%
\mathbf{k},\mathbf{Q}},\gamma^\textbf{p}_{\mathbf{k},\mathbf{Q}}$ or $\widetilde{%
\gamma }^\textbf{p}_{\mathbf{k},\mathbf{Q}}$.

 The vanishing of the secular determinant (\ref{BSZ})  provides the spectrum of the
collective excitations $\omega(\textbf{Q})$. $I^\textbf{p}_{a,b}$
and $J^\textbf{p}_{a,b}$ will display singularities if
$\omega(\textbf{Q})=E_+(\textbf{k}_0;\textbf{p})+E_+(\textbf{Q}-\textbf{k}_0;\textbf{p})$
 at a
particular $\textbf{k}_0$. This means that the superfluid state is
not stable because the Cooper pairs break into two fermionic
excitations.\cite{Com} For fixed $\textbf{p}$ and different
$\textbf{Q}$ the expression
$E_+(\textbf{k};\textbf{p})+E_+(\textbf{Q}-\textbf{k};\textbf{p})$
is bounded from below by the threshold line
$\omega^{th}_{\textbf{p}}(\textbf{Q})=
(min_{\textbf{k}}\left[E_+(\textbf{k};\textbf{p})+
E_+(\textbf{Q}-\textbf{k};\textbf{p})\right]$. Our numerical
calculations in 1D, 2D and 3D show that the threshold line
$\omega^{th}_{\textbf{p}}(\textbf{Q})$ is above the spectrum of the
collective excitations $\omega(\textbf{Q})$, and therefore, the
rotonlike minimum will reach zero before the pair breaking sets in,
i.e. the superfluid state is destabilized due to the Landau
mechanism.
\begin{center}
\begin{figure}[tbp]
\includegraphics[scale=0.95]{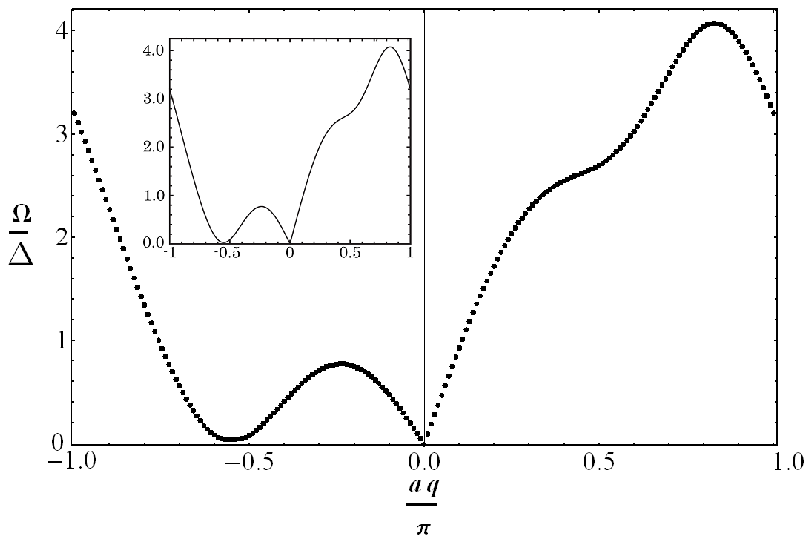}
\caption{Collective-mode spectrum $\Omega(q)$ in 1D optical lattices
for quasimomentum $p=0.21$ calculated by using the secular
determinant (\ref{BSZ}). We set $f=0.5$ and $|U|=2J$. The superfluid
gap and chemical potential are $\Delta=0.419J$ and $\mu=0.655J$. The
insert shows the results from FIG. 4b in Ref. [\onlinecite{Yosh}],
but calculated with $\Delta=0.420J$ and $\mu=0.624J$.
}\end{figure}\end{center}
\begin{center}\begin{figure}[tbp]
\includegraphics[scale=0.95]{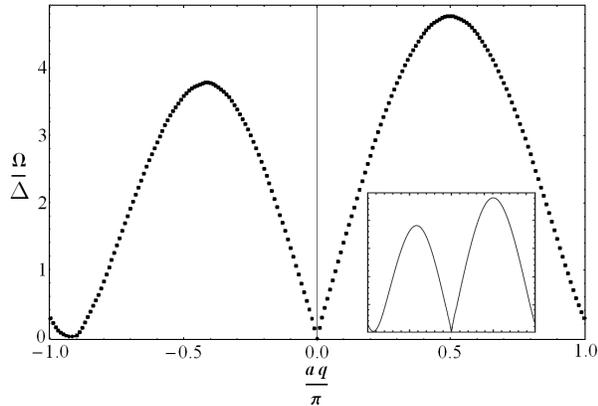}
\caption{Collective-mode spectrum $\Omega(q)$ in 1D optical lattices
for quasimomentum $p=0.12$ calculated by using determinant
(\ref{BSZ}). We set $f=0.9$ and $|U|=2J$. The superfluid gap and
chemical potential are $\Delta=0.351J$ and $\mu=1.697J$. The insert
shows the results from FIG. 6b in Ref. [\onlinecite{Yosh}], but
calculated with $\Delta=0.350J$ and $\mu=1.690J$. }\end{figure}
\end{center}
\begin{center}\begin{figure}[tbp]
\includegraphics[scale=0.95]{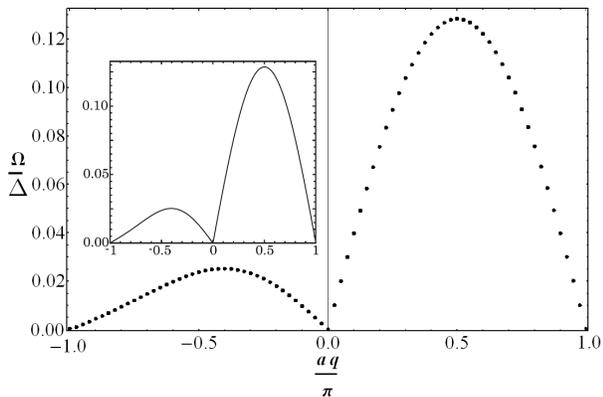}
\caption{Collective-mode spectrum $\Omega(q)$ in 2D optical lattices
for quasimomentum $\textbf{p}=(p,p)$, where $p= 0.6502/\pi\sqrt{2}$
calculated by using the secular determinant (\ref{BSZ}). We set
$f=0.5$ and $|U|=12J$. The superfluid gap and chemical potential are
$\Delta=4.9699J$ and $\mu=0.7455J$. The insert shows the results
from FIG. 9d in Ref. [\onlinecite{Yosh}] with the same $\Delta$ and
$\mu$.}\end{figure}
\end{center}

\begin{center}\begin{figure}[tbp]
\includegraphics[scale=0.9]{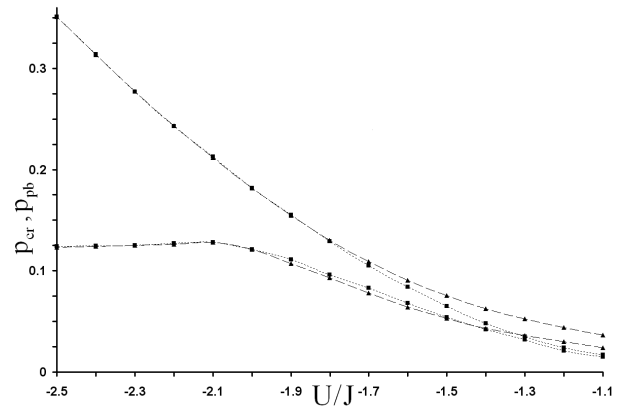}
\caption{Critical  $p_{cr}$ (lower curves) and pair-breaking
$p_{pb}$ (upper curves) quasimomenta in units [$\pi/a$] as a
functions of $U/J$ in 1D lattices and
 $f=0.9$. The dashed curves represent our calculations, while the
dotted curves are the results presented in Fig. 7 in Ref.
[\onlinecite{Yosh}]). }\end{figure}
\end{center}

\section{Comparison with other approximations}
Our numerical calculations show that there is an excellent agreement
between the dispersions obtained by Ganesh et al. in Ref.
 [\onlinecite{Gam}] by using determinant (\ref{Gam}), and by our
 secular determinant (\ref{BSZ}) not only in the case of a stationary lattice
 (see the conclusion section in Ref.
 [\onlinecite{ZK1}]), but in the case of a moving lattice as well. This is
due to the fact that the  terms with index $\widetilde{\gamma}$ in
the secular determinant (\ref{BSZ}) are very small. This is due to
the fact that the  terms with index $\widetilde{\gamma}$ in the
secular determinant (\ref{BSZ}) are very small.

In Fig. 3 and Fig. 4 we have compared the collective mode spectrum
in 1D lattices obtained by the BS approach and by the Green's
function method.\cite{Yosh} It is worth mentioning that in the 1D
case our chemical potentials and gaps obtained from Eqs. (\ref{MO})
are not the same as in Ref. [\onlinecite{Yosh}], and therefore, the
direct comparison between the corresponding dispersion curves is not
correct, but the curves are quite similar. When using the same
values for the chemical potential and the gap, our dispersion curve
is slightly below the corresponding curve obtained in Ref.
[\onlinecite{Yosh}] (see Fig. 5).

In Fig. 6 we have presented the critical $p_{cr}$ (lower curves) and
pair-breaking $p_{pb}$ (upper curves) quasimomenta in 1D lattices as
functions of $U/J$. The Green's function and BS methods provide
similar results for $-2.5<U/J<-1.8$, but in the interval
$-1.7<U/J<-1.1$, both the pair-breaking and the critical
quasimomentum curves differ from the corresponding curves presented
in Fig. 7, Ref. [\onlinecite{Yosh}]. The differences between the
pair-breaking curves are probably due to the different solutions of
gap and number equations, but the differences between the critical
momentum curves are most like due to the different techniques
employed to obtain the collective-mode spectrum. It can be seen that
according to our curves the Landau instability is still the
mechanism which destabilized the superfluid states.

It is interesting to compare the
 analytical results for the speed of sound in a weak-coupling
regime in  1D optical lattices obtained by both approaches. In a
stationary lattice we follow the calculations by Belkhir and
Randeria\cite{BR} to reduce our secular determinant (\ref{BSZ}) to
the following $2\times 2$ determinant:
 \begin{equation}
\left|
\begin{array}{cc}
-\frac{Q^2\alpha(\mu^2-4J\mu+\Delta^2)}{4\Delta^2}-\frac{\omega^2\alpha}{4\Delta^2}
&-\frac{\omega\alpha}{2\Delta}\\
-\frac{\omega\alpha}{2\Delta} & 1-\alpha
\end{array}%
\right|=0,\label{DBR1a}
\end{equation} where $\alpha=|U|/(\pi
\sqrt{4J\mu-\mu^2}$.  The solution $\omega(Q)=cQ$ provides the speed
of sound which in a stationary lattice is
$c_s=\sqrt{1-\alpha}\sqrt{V_F^2-\Delta^2}$, where the Fermi velocity
is $V_F=\sqrt{4J\mu-\mu^2}$.  Using the same steps as in a
stationary lattice, we have obtained the following $2\times 2$
 determinant in 1D moving optical lattices:
\begin{equation}
\left|
\begin{array}{cc}
a_{11}
&-\frac{\omega\alpha}{2\Delta}+\frac{(2J-\mu)Q\alpha\tan(p)}{2\Delta}\\
-\frac{\omega\alpha}{2\Delta}+\frac{(2J-\mu)Q\alpha\tan(p)}{2\Delta}
& 1-\alpha
\end{array}%
\right|=0.\label{DBR1}
\end{equation}
where $\alpha=|U|/(\pi \sqrt{4J^2\cos^2(p)-(2J-\mu)^2})$ and
\begin{widetext}
\begin{equation}a_{11}= -\frac{Q^2\alpha(\mu^2-4J\mu+\Delta^2)}{4\Delta^2\cos^2(p)}
-\frac{\omega^2\alpha}{4\Delta^2}
+\frac{(2J-\mu)Q\alpha\omega\tan(p)}{2\Delta^2}-\frac{1}{4}Q^2\tan^2(p)
-\frac{[2J^2-\Delta+4J^2\cos(2p)]Q^2\alpha\tan^2(p)}{2\Delta^2}.
\label{A11}\end{equation} The last determinant provides the
following expression for the phononlike dispersion in the
long-wavelength limit:
\begin{equation}\begin{split}&c_s=(2J-\mu)\tan(p)
+\sqrt{1-\alpha}\sqrt{V^2_F-
\Delta^2\left[1+\frac{1-\alpha}{\alpha}\tan^2(p)\right]},\\&
V_F=\sqrt{4J^2\cos^2(p)-(2J-\mu)^2+4J^2\left[2-3\cos^2(p)\right]\tan^2(p)}.
\label{1D}\end{split}\end{equation}  The speed of sound obtained in
Ref. [\onlinecite{Yosh}] is
\begin{equation}\begin{split}&c_s=(2J-\mu)\tan(p)
+\sqrt{1-\alpha}\sqrt{V^2_F-
\Delta^2\left[1+\frac{1-\alpha}{\alpha^2}\tan^2(p)\right]},\\&
V_F=\sqrt{4J^2\cos^2(p)-(2J-\mu)^2}.
\label{1DY}\end{split}\end{equation}
\end{widetext}
It should be mentioned that in the case when $\Delta/J<1$, the term
$4J^2\left[2-3\cos^2(p)\right]\tan^2(p)$  is more important than the
term $\Delta^2(1-\alpha)\tan^2(p)/\alpha$ in (\ref{1D}) and the term
$\Delta^2(1-\alpha)\tan^2(p)/\alpha^2$ in (\ref{1DY}).
\section{Summary}
In conclusion, we have derived the BS equation in GRPA for the
collective-mode spectrum of superfluid Fermi gases in moving optical
lattices assuming that the system is described by the attractive
Hubbard model. We have compared collective excitation spectrum
obtained by the BS approach with the corresponding spectrum obtained
by applying the Green's function formalism and by the perturbation
theory. The BS results are in an excellent agreement with the
results obtained by perturbation theory, while the  Green's function
formalism provides slightly different results.

\end{document}